\documentclass[10pt,final,journal,twocolumn]{IEEEtran}
\ifCLASSINFOpdf
\else
   \usepackage[dvips]{graphicx}
\fi
\usepackage[cmex10]{amsmath}
  \usepackage{cite}
\usepackage{cite}
\usepackage[cmex10]{amsmath}
\usepackage{amsthm}
\usepackage{algorithmic}
\usepackage{amsfonts}
\usepackage{stfloats}
\usepackage{multicol,multienum}
\usepackage[dvips]{graphicx}

  \usepackage{cite}
\usepackage{cite}
\usepackage[cmex10]{amsmath}
\usepackage{amsthm}
\usepackage{algorithmic}
\usepackage{amsfonts}
\usepackage{stfloats}
\usepackage{multicol,multienum}
\usepackage[dvips]{graphicx}
\usepackage{cite} 
\usepackage{url}  
\usepackage{ifthen}  
\usepackage{multicol}   
\usepackage[utf8]{inputenc} 
\usepackage{amsfonts}
\usepackage{amssymb}

\begin{document}
%
\title{When Does Relay Transmission Give a More Secure Connection in Wireless Ad Hoc Networks?}
%
%
%

\author{Chunxiao Cai,
        Yueming Cai,
         Xiangyun Zhou,
         Weiwei Yang,
        and Wendong Yang
\thanks{This work is supported by the Project of Natural Science Foundations of China (No. 61371122, No. 61001107 and No. 61301163).}
\thanks{Chunxiao Cai, Yueming Cai, Weiwei Yang, and Wendong Yang are all with the College of Communications Engineering, PLA University of Science and Technology, Nanjing 210007, China (email: caichunxiao1007@126.com, caiym@vip.sina.com,  yww\_1010@yahoo.com.cn, and ywd1110@163.com).

  Xiangyun Zhou is with Research School of Engineering, The Australian National University, Australia. (email: xiangyun.zhou@anu.edu.au). }}

\maketitle

\begin{abstract}

Relay transmission can enhance coverage and throughput, while it can be vulnerable to eavesdropping attacks due to the additional transmission of the source message at the relay. Thus, whether or not one should use relay transmission for secure communication is an interesting and important problem. In this paper, we consider the transmission of a confidential message from a source to a destination in a decentralized wireless network in the presence of randomly distributed eavesdroppers. The source-destination pair can be potentially assisted by randomly distributed relays. For an arbitrary relay, we derive exact expressions of secure connection probability for both colluding and non-colluding eavesdroppers. We further obtain lower bound expressions on the secure connection probability, which are accurate when the eavesdropper density is small. By utilizing these lower bound expressions, we propose a relay selection strategy to improve the secure connection probability. By analytically comparing the secure connection probability for direct transmission and relay transmission, we address the important problem of whether or not to relay and discuss the condition for relay transmission in terms of the relay density and source-destination distance. These analytical results are accurate in the small eavesdropper density regime.
\end{abstract}


%
\IEEEpeerreviewmaketitle

\section{Introduction}
%
%
%
%
\subsection{Background}
\IEEEPARstart{I}{nformation}
 security of wireless communications has taken on an increasingly important role as these networks continue to flourish worldwide.
Communications over wireless networks are particularly vulnerable to eavesdropping attacks due to the broadcast nature of the wireless medium. To protect confidential message transmission, physical layer security [1] has been developed as a promising mechanism which provides the protection at the physical layer by exploiting the random and noisy nature of the wireless propagation channels. A key feature of physical layer security is that the level of security provided strongly depends on the amount of information that the legitimate users know about the eavesdroppers. In particular, perfect secrecy may not be always achievable if the channel state information (CSI) of the eavesdropper is not perfectly known. For this practical scenario, Bloch $et~al.$ studied the secure outage probability assuming no instantaneous CSI of the eavesdropper at the legitimate transmitter [2]. Considering that the locations of passive eavesdroppers are uncertain, e.g., in military communication networks, there are a few recent studies which modeled the distribution of eavesdroppers as a homogenous Poisson point process (PPP) and studied the connectivity properties of such networks [3]-[7]. However, most of these papers mainly focused on single-hop transmissions, while multi-hop transmission was only studied from a percolation point of view.

Due to their ability to extend coverage in wireless systems, relay transmission technologies have received much attention. When secure communication is required, however, relay transmission may be vulnerable to eavesdropping, because the confidential information is broadcasted twice, i.e., by the source and the relay. Thus, it is very interesting to study the physical-layer security in relay networks. In [8], Lai  and  El Gamal studied the four-node (source, destination, relay, eavesdropper) secure communication system and considered several relay strategies, such as decode-and-forward (DF) and noise-forwarding (NF). In [9], Jeong and Kim analyzed the power allocation problem for secure rate maximization in a multi-carrier decode-and-forward relay system. In [10], L. Dong $et~al.$ considered a scenario in which a source communicates with a destination with the help of multiple relays in the presence of one or more eavesdroppers and used beamforming to improve the security. Krikidis $et~al.$ studied relay and jammer selection in relay systems with secrecy constraints [11]. In [12],  Chen $et~al.$ investigated joint relay and jammer selection in two-way relay networks with security constraints. Furthermore, our prior work in [13] studied the problem of secure connectivity against colluding eavesdroppers where the source-destination pair is assisted by a single randomize-and-forward (RaF) relay defined in  [14] [15].  All above papers studied different ways of using relay transmission to improve secrecy performance of relay networks without the direct link between the source and destination. However, these papers did not directly answer the basic question: With the purpose of achieving a good secrecy performance, shall we use the relay if there is already a direct link between the source and destination?
\subsection{Approach and Contributions}
In this paper, we try to address the question of whether or not to use relay transmission in wireless ad hoc networks from a secure connectivity perspective. We consider that the locations of both the potential relays and malicious eavesdroppers are random following homogenous PPPs. The widely-used RaF relaying strategy is assumed, which was specifically designed for secure transmissions [14] [15]. Both the cases of colluding and non-colluding eavesdroppers are considered, where the former represents a worst-case scenario and the latter is a commonly assumed scenario in decentralized wireless networks.

Firstly, assuming that a relay at an arbitrary location is already chosen, we derive the exact expression of secure connection probability for relay transmission. Then, a lower bound on the secure connection probability is provided in order to further study the optimal relay selection strategy. We also derive the secure connection probability for direct transmission from the source to the destination. Having the secrecy performance of both relay transmission and direct transmission, we provide an answer to whether or not to use relay transmission. In particular, we give an analytical expression on the condition for relay transmission in terms of the relay density and the distance between the source and destination, for a given target secure connection probability. The analytical results are very accurate in the small eavesdropper density regime.

The rest of this paper is organized as follows. In Section II,
we study the system model.
Section III obtains the performance analysis for colluding eavesdroppers. In
Section IV, we analyze the performance analysis for non-colluding eavesdroppers. Section V studies the condition for relay transmission.
Concluding remarks are
provided in Section VI.

\section{System Model}
As shown in Fig.1, we consider a relay network consisting of one source ($S$), several relays ($R_l ,l = 1,2,...$), one destination ($D$), and several eavesdroppers ($E_j ,j = 1,2,...$). All the nodes are equiped with one antenna. The distance between the source and destination is equal to  $d_{SD}$. The distributions of relays and eavesdroppers are  homogenous PPPs   $ \Phi _R$ and $ \Phi _E$  with density $\lambda _R$  and  $\lambda _E$, respectively.
   In this system, all the transmitters  transmit with the same power. Then we can obtain the instantaneous signal-to-noise ratio (SNR) at the relays, destination, and eavesdroppers as
\begin{equation}\label{1}
SNR_{nm}  = \varepsilon h_{nm} d_{nm}^{ - \alpha }
\end{equation}
where $\varepsilon$   is the transmit SNR,  $h_{nm}$  and $d_{mn} \!\! = \!\!\left\| {x_m \! -\! x_n } \right\|$    mean the small-scale fading and the Euclidean distance between node  $n$  and node  $m$, respectively, and $x_n$ is the location of node $n$. We assume that  $h_{nm}$  follows the exponential distribution with unit mean, and  $d_{nm}^{ - \alpha }$ denotes the large-scale fading with path-loss exponent  $\alpha  > 2$. We assume the source has the instantaneous CSI of the links from the source to the relays and from the source to the destination, and the relays have the instantaneous CSI of the links from the relays to the destination. We assume the source and the relays do not have the instantaneous CSI of the links from the source to the eavesdroppers and from the relays to the eavesdroppers, respectively. Furthermore, receiver side CSI is always assumed.

In this paper, the source performs relay selection and decides whether a relay is needed. We assume that  $d_{SD}$, and the node densities, i.e.,  $\lambda _E$ and  $\lambda _R$, are known network parameters. The relays can measure their distances to both the source and the destination, then report the distances to the source.

To facilitate the analysis, a polar coordinate system
is set up in which the  source and destination locates at $\left( {\frac{{d_{SD} }}{2},0} \right)$  and  $\left( {\frac{{d_{SD} }}{2},\pi } \right)$, respectively.  In a polar coordinate system, for a  relay at  $x_{R_l }   = \!\left( {r,\theta } \right)$, define two auxiliary functions to represent the distances from the arbitrary relay to the source and the destination respectively:  $d_{SR_l }^{}  \!=\! \left\| {x_S \! - \!x_{R_l } } \right\|\! = \!\!\sqrt {r^2  +\! \frac{{d_{SD}^2 }}{4} -\! rd_{SD}^{} \cos \theta } $, and  $d_{R_l D}^{}  =\! \left\| {x_{R_l }\!  - \!x_D } \right\| \!\!=\!\! \sqrt {r^2 \! +\!\frac{{d_{SD}^2 }}{4}+\! rd_{SD}^{} \cos \theta }$.

\begin{figure}
\begin{center}
  \includegraphics[width=3.0in,angle=0]{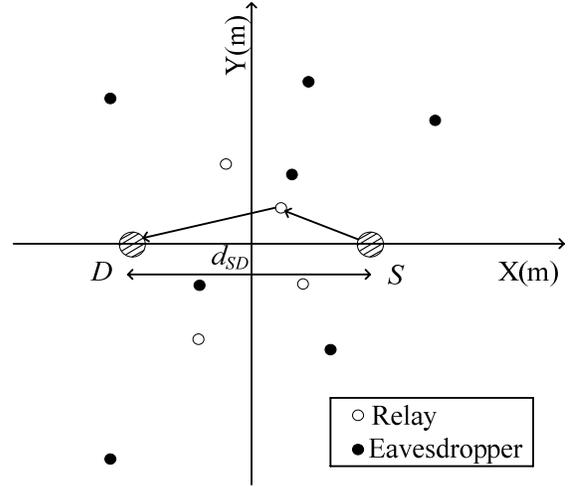}\\
  \caption{System Model.}
\end{center}
\end{figure}

\section{Performance Analysis for Colluding Eavesdroppers}

 There are two types of eavesdroppers depending on whether the eavesdroppers combine the received
information: colluding
eavesdroppers and non-colluding eavesdroppers [7].
The colluding eavesdroppers case is the worst
scenario from the secure communication viewpoint. It means that all the eavesdroppers can exchange and combine the received
information to decode the secret message. On the other hand,
the non-colluding eavesdroppers case means that
the eavesdroppers are not allowed to collude, and secure performance is determined with the strongest
received signal from the transmitter.
In this section, we study the  secure connection probability for colluding  eavesdroppers, and obtain the exact expressions of secure connection probability for the  direct transmission and relay transmission assuming an arbitrary relay, respectively.
 Then the lower bound for colluding  eavesdroppers is obtained, and the lower bound gives accurate approximation of the exact performance when the  eavesdropper density is small. Using the lower bound, we find that the optimum relay is the nearest one to the midpoint between the source and destination, and get the  lower bound expression for relay selection.

For colluding eavesdroppers case, the combined received SNR at eavesdroppers from the source and the relay  node $R_l$  can be respectively written as [7]
\begin{equation}\label{2}
\varepsilon I_S  = \sum\limits_{E_j  \in \Phi _E } {\varepsilon h_{SE_j } d_{SE_j }^{ - \alpha } }
\end{equation}
\begin{equation}\label{3}
\varepsilon I_{R_l }  = \sum\limits_{E_j  \in \Phi _E } {\varepsilon h_{R_l E_j } d_{R_l E_j }^{ - \alpha } }
\end{equation}

\subsection{Direct Transmission}
From a connectivity
point of view, a secure connection from the source
to the destination is possible if the secrecy rate is
positive [7].  The secure connection probability for direct transmission can be defined as
\begin{equation}\label{4}
\begin{array}{l}
 P_{C\_DT}  = P\left( {\log _2 \big( {\frac{{1 + \varepsilon h_{SD} d_{SD}^{ - \alpha } }}{{1 + \varepsilon I_S }}} \big) > 0} \right)\\
 ~~~~~~~~~\!{\rm{     }}=  {\mathop{P}\nolimits} \left( {\frac{{\varepsilon h_{SD} d_{SD}^{ - \alpha } }}{{\varepsilon I_S }} > 1} \right) \\
 ~~~~~~~~~\!{\rm{     }} = E_{I_S } \left\{ {\exp \left[ { - I_S d_{SD}^\alpha  } \right]} \right\} \\
  ~~~~~~~~~\!= L_{I_S } \left( {d_{SD}^\alpha  } \right) \\
 \end{array}
\end{equation}
where  $E_{I_S } \left\{  \cdot  \right\}$ and $L_{I_S } \left(  \cdot  \right)$  is expected value and Laplace transform of  $I_S$, respectively. Then we can get  [16]
\begin{equation}\label{5}
\begin{array}{l}
 L_{I_S } \left( {d_{SD}^\alpha  } \right) \\
  = E_{h_{SE_j } ,\Phi _E } \left\{ {\exp \left[ { - \sum\limits_{E_j  \in \Phi _E } {d_{SD}^\alpha  } h_{SE_j } \left\| {x_S  - x_{E_j } } \right\|^{ - \alpha } } \right]} \right\} \\
  = E_{h_{SE_j } ,\Phi _E } \left\{ {\prod\limits_{E_j  \in \Phi _E } {\exp \left[ { - d_{SD}^\alpha  h_{SE_j } \left\| {x_S  - x_{E_2 } } \right\|^{ - \alpha } } \right]} } \right\} \\
 \mathop  = \limits^{\left( \omega  \right)}\!\! E_{\Phi _E } \left\{ {\prod\limits_{E_j  \in \Phi _E }\!\! {E_{h_{SE_j } }\! \!\left\{ {\exp \left[ { - d_{SD}^\alpha  h_{SE_j } \left\| {x_S  - x_{E_j } } \right\|^{ - \alpha } } \right]} \right\}} } \right\} \\
  = E_{\Phi _E } \left\{ {\prod\limits_{E_j  \in \Phi _E } {\frac{1}{{1 + d_{SD}^\alpha  \left\| {x_S  - x_{E_j } } \right\|^{ - \alpha } }}} } \right\} \\
 \end{array}
\end{equation}
where $\omega $ means that ${h_{SE_j } }$ is independent and identically distributed fading and  independent of $\Phi _E$, thus the expectation over ${h_{SE_j} }$ can be moved inside the product. For a homogenous PPP, the generating function is given by [17]
\begin{equation}\label{6}
 E_{\Phi _E }  \Big\{ {\prod\limits_{E_j  \in \Phi _E } {f\left( {x_{E_j } } \right)} } \Big\} = \exp \left[ { - \lambda _E \int_{\mathbb{R}^2 } {\Big( {1 - f\left( {x_{E_j } } \right)} \Big)dx_{E_j } } } \right]
 \end{equation}

 Then (5) can be derived as
\begin{equation}\label{7}
\begin{array}{l}
 P_{C\_DT}  = \exp \left[ { - \lambda _E \int_{\mathbb{R}^2 } { 1 - \frac{1}{{1 + d_{SD}^\alpha  \left\| {x_S  - x_{E_j } } \right\|^{ - \alpha } }}dx_{E_j } } } \right] \\
 ~~~~~~~~{\rm{              }} = \exp \left[ { - Ad_{SD}^2 } \right] \\
 \end{array}
\end{equation}
where  $A \!= \!\frac{{2\pi \lambda _E }}{\alpha }\Gamma \!\left( {\frac{2}{\alpha }} \right)\Gamma \!\left( {1 \!-\! \frac{2}{\alpha }} \right)$, $\Gamma \!\left(  \cdot  \right)$  is the gamma function.

\subsection{Relay Transmission}
The communication protocol is a hybrid mode that allows adaptive switching between
direct transmission and relay transmission. In this
paper, however, we consider a switching criterion with priority
on the use of a direct transmission. Then, the relay transmission
is only used when the direct path does not satisfy the target
security requirement.
For relay transmission,  we use RaF strategy. According to [14],  RaF means that the source and relay use independent randomization signal in each hop.
The communication is divided into two slots. In the first time slot, the source sends data to the relays. In the second time slot, only the selected relay sends data to the destination.  In [14], it has been shown  that under the RaF strategy, securing each individual hop is sufficient for securing the end-to-end transmission, so the message is secured if the two hops are both secured.

For an arbitrary relay $R_l$, the message is secure only if both the $S \to R_l$  link and the $R_l \to D$ link are secure. Thus, the secure connection probability can be obtained as
\begin{equation}\label{8}
\begin{array}{l}
 P_{C\_R_l }  \\
  = {\mathop{ P}\nolimits} \left( {\frac{1}{2}\log _2 \!\left( {\frac{{1 + \varepsilon h_{SR_l } d_{SR_l }^{ - \alpha } }}{{1 + \varepsilon I_S }}} \!\right) > 0,\!\frac{1}{2}\log _2\! \left( {\frac{{1 + \varepsilon h_{SR_l } d_{R_l D}^{ - \alpha } }}{{1 + \varepsilon I_{R_l } }}}\! \right)\! >\! 0} \right) \\
  = {\mathop{ P}\nolimits} \left( {\frac{{h_{SR_l } d_{SR_l }^{ - \alpha } }}{{I_S }} > 1,\frac{{h_{SR_l } d_{R_l D}^{ - \alpha } }}{{I_{R_l } }} > 1} \right) \\
  = E_{I_S ,I_{R_l } } \left( {\exp \left[ { - I_S d_{SR_l }^\alpha   - I_{R_l } d_{R_l D}^\alpha  } \right]} \right) \\
  = L_{I_S ,I_{R_l } } \left( {d_{SR_l }^\alpha  ,d_{R_l D}^\alpha  } \right) \\
 \end{array}
\end{equation}
where $L_{I_S ,I_{R_l } } \left( { \cdot , \cdot } \right)$  is the joint Laplace transform of  $I_S$ and  $I_{R_l }$ [18], and it can be calculated as (9), shown at the top of the next page.
\setcounter{equation}{8}
\begin{figure*}[bt!]
\begin{equation}\label{9}
\begin{array}{l}
 L_{I_S ,I_{R_l } } \left( {d_{SR_l }^\alpha \! ,d_{R_l D}^\alpha  } \right)\!\! =\!\! E_{h_{SE_j ,} h_{R_l E_j } ,\Phi _E } \Big\{ {\exp \Big[ {\! -\! \sum\limits_{E_j  \in \Phi _E } {\left( {d_{SR_l }^\alpha  h_{SE_j } \left\| {x_S \! - \!\!x_{E_j } } \right\|^{ -\! \alpha } \! \!+ \!d_{R_l D}^\alpha  h_{R_l E_j } \left\| {x_{R_l } \!\! -\! x_{E_j } } \right\|^{ -\! \alpha } } \right)} } \Big]} \Big\} \\
 ~~~~~~~~~~~~~~~~~~~~~~~{\rm{         }} = E_{\Phi _E } \Big\{ {\prod\limits_{E_j  \in \Phi _E } {\Big( {\frac{1}{{1 + d_{SR_l }^\alpha  \left\| {x_S  - x_{E_j } } \right\|^{ - \alpha } }} \cdot \frac{1}{{1 + d_{R_l D}^\alpha  \left\| {x_{R_l }  - x_{E_j } } \right\|^{ - \alpha } }}} \Big)} } \Big\} \\
 \end{array}
\end{equation}
\rule{\linewidth}{1pt}
\end{figure*}


Using (6), then,  the secure connection probability for  an arbitrary relay can be written as
\begin{equation}\label{10}
P_{C\_R_l }  = \exp \left[{ - \left( {Ad_{SR_l }^2  + Ad_{R_l D}^2  - \lambda _E f\left( {d_{SR_l }^\alpha  ,d_{R_l D}^\alpha  } \right)} \right)} \right]
\end{equation}

In (10), $f\left( {d_{SR_l }^\alpha  ,d_{R_l D}^\alpha  } \right)$ can be obtained as
 \begin{equation}\label{11}
\begin{array}{l}
 f\left( {d_{SR_l }^\alpha  ,d_{R_l D}^\alpha  } \right) \\
  = \int_{\mathbb{R}^2 } {\frac{1}{{\left( {1 + d_{SR_l }^{ - \alpha } \left\| {x_S  - x_{E_j } } \right\|^\alpha  } \right)\left( {1 + d_{R_l D}^{ - \alpha } \left\| {x_{R_l }  - x_{E_j } } \right\|^\alpha  } \right)}}} dx_{E_j }  \\
 \end{array}
\end{equation}

Notice that  $f\left( {d_{SR_l }^\alpha  ,d_{R_l D}^\alpha  } \right)$  in (11) accounts for the statistical dependence between the received SNRs  at two different locations.

From (11), we can find that the function  $f\left( { \cdot , \cdot } \right)$ can not be computed in a closed form.
From (10) and (11), we can also find that when $\lambda _E  \to 0$, $\exp \left[ { \lambda _E f\left( {d_{SR_l }^\alpha  ,d_{R_l D}^\alpha  } \right)} \right] \to 1$.
Then, we can get the lower bound of (10). It shows that $f\left( {d_{SR_l }^\alpha  ,d_{R_l D}^\alpha  } \right)$  is non-negative, thus, the lower bound of the secure connection probability can be defined as
\begin{equation}\label{12}
P_{C\_R_l \_lower} {\rm{ }} = \exp \left[{ - \left( {Ad_{SR_l }^2  + Ad_{R_l D}^2 } \right)} \right]
\end{equation}

Substituting  $d_{SR_l}^{}  = \sqrt {r^2  + \frac{{d_{SD}^2 }}{4} - rd_{SD}^{} \cos \theta }$, and  $d_{R_lD}^{}  = \sqrt {r^2  + \frac{{d_{SD}^2 }}{4} + rd_{SD}^{} \cos \theta }$ into (12), (12) can be written as
\begin{equation}\label{13}
P_{C\_R_l \_lower} {\rm{ }} = \exp \left[ { - \frac{{Ad_{SD}^2 }}{2} - 2Ar^2 } \right]
\end{equation}

From (10) and (12), we can get
\begin{equation}\label{14}
\begin{array}{l}
\frac{{P_{C\_R_l } }}{{P_{C\_R_l \_lower} }}  = \exp \left[ {\lambda _E f\left( {d_{SR_l }^\alpha  ,d_{R_l D}^\alpha  } \right)} \right]\\
~~~~~~~~~~~~~{\rm{         }} < \exp \left[ {\min \left\{ {Ad_{SR_l }^2 ,Ad_{R_l D}^2 } \right\}} \right]
 \end{array}
\end{equation}

It shows that when $\lambda _E  \to 0$, $\frac{{P_{C\_R_l } }}{{P_{C\_R_l \_lower} }} \to 1$.
Fig.2 shows secure connection probability for colluding eavesdroppers as a function of  $\lambda _E$, where $\left( { \cdot , \cdot } \right)$  means the relay location.  It is shown that the secure connection probability decreases with the increase of $\lambda _E$.
Through the gap between the lower bound and the exact value becomes larger with the increase of $\lambda _E$, we can find that the result obtained by using the lower bound gives accurate approximation of the exact performance when the distribution of eavesdroppers is sparse.
Then we can use (13) to analyze the secure connection probability instead of (10) in the small eavesdropper density regime.

\begin{figure}
\begin{center}
  \includegraphics[width=3.4in,angle=0]{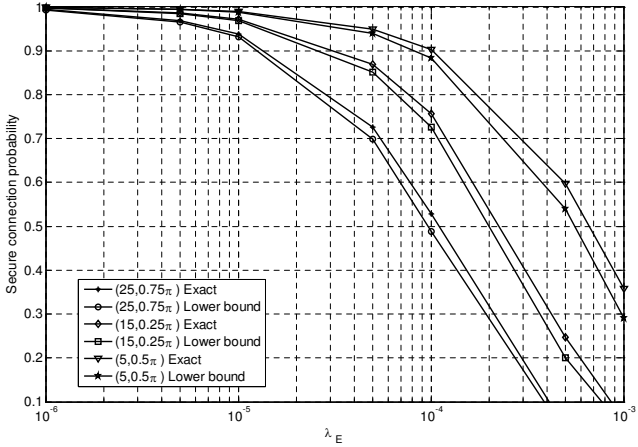}\\
  \caption{Secure connection probability for colluding eavesdroppers as a function of  $\lambda _E$
  when  $\alpha  = 4$, $d_{SD}  = 20m$.}
\end{center}
\end{figure}

\subsection{Relay Selection}

From Fig.2, we  can also find that the location of the arbitrary relay has influence on the secure connection probability. Then, we can improve the secure connection probability through a relay selection strategy.
It shows in (13) that when $r  \to 0$, $P_{C\_R_l \_lower} {\rm{ }}  \to \exp \big[ { - \frac{{Ad_{SD}^2 }}{2} } \big]$. Thus,
we can find the relay which is nearest to the midpoint between the source and destination has the better secure connectivity probability. Thus the index of the best relay can be shown as
\begin{equation}\label{15}
J = \arg \mathop {\min }\limits_{R_l  \in \Phi _R } \left\{ {\left\| {x_{R_l } } \right\| } \right\}
\end{equation}

The probability distribution of $\left\| {x_{R_J } } \right\|$  is [19]
\begin{equation}\label{16}
P_{\left\| {x_{R_J } } \right\|} \left( r \right) = \lambda _R \exp \left[ { - \lambda _R \pi r^2 } \right]
\end{equation}

From the homogeneity of PPP, the angle of $\left\| {x_{R_J } } \right\|$ should be a uniform random variable in $\left. {\left[ {0,2\pi } \right.} \right)$, independent of its norm.
From (13) and (16), the lower bound of the secure connection probability can be expressed as
\begin{equation}\label{17}
\begin{array}{l}
 P_{C\_R_J \_lower}  \\
  = \int_{\mathbb{R}^2 } {\exp \left[ { - \left( {\frac{{Ad_{SD}^2 }}{2} + 2Ar^2 } \right)} \right]} \lambda _R \exp \left[ { - \lambda _R \pi r^2 } \right]dx_{R_J }  \\
  = \exp \left[ { - \frac{{Ad_{SD}^2 }}{2}} \!\right]\int_0^{2\pi } {\int_0^{ + \infty } {\lambda _R \exp \left[ { - \left( {2A + \lambda _R \pi } \right)r^2 } \right]} } rdrd\theta  \\
  = \frac{{\lambda _R \pi }}{{2A + \lambda _R \pi }}\exp \left[ { - \frac{{Ad_{SD}^2 }}{2}} \right] \\
 \end{array}
 \end{equation}


\section{Performance Analysis for Non-Colluding Eavesdroppers}

We will investigate  the case of non-colluding eavesdroppers. Assume eavesdroppers can not be allowed to collude and exchange information. The secure performance is determined with the strongest
received signal from the transmitter. In this section, we obtain the exact expressions of secure connection probability for the  direct transmission and relay transmission assuming an arbitrary relay, respectively.
 Then we analyze the lower bound for non-colluding  eavesdroppers and find that  the lower bound is close to the exact performance when the  eavesdropper density is small. Using the same optimum relay selection strategy with colluding  eavesdroppers, we obtain the lower bound expressions for relay selection.

For non-colluding eavesdroppers case, the received SNR at eavesdroppers from the source and  relay  node $R_l$  can be respectively written as
\begin{equation}\label{18}
\varepsilon U_S  =  \mathop {\max }\limits_{E_j  \in \Phi _E } \Big\{ {\varepsilon h_{SE_j } d_{SE_j }^{ - \alpha } } \Big\}
\end{equation}
\begin{equation}\label{19}
\varepsilon U_{R_l }  =  \mathop {\max }\limits_{E_j  \in \Phi _E } \Big\{ {\varepsilon h_{R_l E_j } d_{R_l E_j }^{ - \alpha } } \Big\}
\end{equation}

\subsection{Direct  Transmission}

Similar to the case of colluding eavesdroppers,
the secure  connection probability for direct transmission can be defined as [7]
\begin{equation}\label{20}
\begin{array}{l}
 P_{C\_DT}  \\
  = {\mathop{ P}\nolimits} \left( {\log _2 \left( {\frac{{1 + \varepsilon h_{SD} d_{SD}^{ - \alpha } }}{{1 + \varepsilon U_S }}} \right) > 0} \right) \\
  = P \left( {\frac{{h_{SD} d_{SD}^{ - \alpha } }}{{ \mathop {\max }\limits_{E_j  \in \Phi _E } \left\{ {h_{SD} \left\| {x_S  - x_{E_j } } \right\|^{ - \alpha } } \right\}}} > 1} \right) \\
  = E_{ {h_{SD} }  ,\Phi _E }\! \left\{ {\prod\limits_{E_j  \in \Phi _E } {P\left( {\frac{{h_{SD} d_{SD}^{ - \alpha } }}{{h_{SD} \left\| {x_S  - x_{E_j } } \right\|^{ - \alpha } }} > 1\left| {h_{SD} ,\Phi _E } \right.} \right)} } \right\} \\
 \end{array}
\end{equation}

Using (6), (20) can be written as
\begin{equation}\label{21}
\begin{array}{l}
 P_{C\_DT}  \\
  = E_{h_{SD} }\! \left\{\!\! {\exp \!\!\left[ { - \lambda _E \!\int_{\mathbb{R}^2 } \!\!{P\!\left( {\frac{{h_{SD} d_{SD}^{ - \alpha } }}{{h_{SE_j } \left\| {x_S \! - x_{E_j } } \right\|^{ - \alpha } }}\! <\! 1\left| {h_{SD} } \!\right.} \!\right)} dx_{E_j } }\! \right]}\! \right\} \\
  = E_{h_{SD} } \left\{ {\exp \left[ { - 2\pi \lambda _E \int_0^{ + \infty } {\exp \left[ { - h_{SD} d_{SD}^{ - \alpha } r^\alpha  } \right]rdr} } \right]} \right\} \\
 \end{array}
\end{equation}

Using [20,3.381.10.4], (21) can be calculated as
\begin{equation}\label{22}
P_{C\_DT}  = E_{h_{SD}^{} } \left\{ {\exp \left[ { - \lambda _E \pi d_{SD}^2 \Gamma \left( {1 + \frac{2}{\alpha }} \right)h_{SD}^{ - \frac{2}{\alpha }} } \right]} \right\}
\end{equation}

Using Jensen$'$s inequality, the lower bound of secure  connection probability can be obtained as
\begin{equation}\label{23}
\begin{array}{l}
 P_{C\_DT}  > \exp \left[ { - \lambda _E \pi d_{SD}^2 \Gamma \left( {1 + \frac{2}{\alpha }} \right)E_{h_{SD}^{} } \left\{ {h_{SD}^{ - \frac{2}{\alpha }} } \right\}} \right] \\
  ~~~~~~~~= \exp \left[ { - Ad_{SD}^2 } \right] \\
 \end{array}
\end{equation}

We can find that (23) is the same as (7), and it means that the secure connection probability for colluding eavesdroppers is worse than that of non-colluding eavesdroppers.

\subsection{Relay Transmission}

 Similar to the case of colluding eavesdroppers, the secure connection probability   for an arbitrary relay can be written  as (24), shown at the top of the next page.
 \setcounter{equation}{23}
\begin{figure*}[bt!]
\begin{equation}\label{24}
\begin{array}{l}
P_{C\_R_l }    = P\left( {\frac{1}{2}\log _2 \big( {\frac{{1 + \varepsilon h_{SR_l } d_{SR_l }^{ - \alpha } }}{{1 + \varepsilon U_S }}} \big) > 0,\frac{1}{2}\log _2 \big( {\frac{{1 + \varepsilon h_{R_l D} d_{R_l D}^{ - \alpha } }}{{1 + \varepsilon U_{R_l } }}} \big) > 0} \right) \\
 ~~~~~~~ = P\Big( {\frac{{\varepsilon h_{SR_l } d_{SR_l }^{ - \alpha } }}{{\mathop { \max }\limits_{E_j  \in \Phi _E } \left\{ {\varepsilon h_{SE_j } \left\| {x_S  - x_{E_j } } \right\|^{ - \alpha } } \right\}}} > 1,\frac{{\varepsilon h_{R_l D} d_{R_l D}^{ - \alpha } }}{{\mathop { \max }\limits_{E_j  \in \Phi _E } \left\{ {\varepsilon h_{R_l E_j } \left\| {x_{R_l }  - x_{E_j } } \right\|^{ - \alpha } } \right\}}} > 1} \Big) \\
~~~~~~~=E_{h_{SR_l } , h_{R_l D} ,\Phi _E } \left\{ {\prod\limits_{E_j  \in \Phi _E } {P\left( {\frac{{h_{SR_l } d_{SR_l }^{ - \alpha } }}{{h_{SE_j } \left\| {x_S  - x_{E_j } } \right\|^{ - \alpha } }} > 1,\frac{{h_{R_l D } d_{R_l D}^{ - \alpha } }}{{h_{R_l E_j} \left\| {x_{R_l }  - x_{E_j } } \right\|^{ - \alpha } }} > 1\left| {h_{SR_l }, h_{R_l D} ,\Phi _E} \right.} \right)} } \right\}
 \end{array}
\end{equation}
\rule{\linewidth}{1pt}\end{figure*}

Using (6), (24) can be derived as (25), shown at the top of the next page.
 \setcounter{equation}{24}
\begin{figure*}[bt!]
\begin{equation}\label{25}
P_{C\_R_l } \!\! =\!\! E_{h_{SR_l } ,h_{R_l D}  } \left\{ {\exp \left[ {\! - \pi \lambda _E \left( {d_{SR_l }^2 \Gamma \Big( {1 \!+ \!\frac{2}{\alpha }} \Big)h_{SR_l }^{ \!- \frac{2}{\alpha }} \! +\! d_{R_l D}^2 \Gamma \Big( {1 \!+ \!\frac{2}{\alpha }} \Big)h_{R_l D}^{\! - \frac{2}{\alpha }} } \right)\! + \lambda _E g\left( {d_{SE_j }^\alpha  ,d_{R_l E_j }^\alpha  } \right)} \right]} \right\}
\end{equation}
\rule{\linewidth}{1pt}\end{figure*}

In (25), $g\left( {d_{SR_l }^\alpha  ,d_{R_l D }^\alpha  } \right)$ can be expressed as (26), written at the top of the next page.
 \setcounter{equation}{25}
\begin{figure*}[bt!]
\begin{equation}\label{26}
g\left( {d_{\!SE_j }^\alpha \! ,\!d_{R_l E_j }^\alpha  } \right) = \int_{\mathbb{\mathbb{R}}^2 } {\exp \left[ { - h_{SR_l } d_{SR_l }^{ - \alpha } \left\| {x_S  - x_{E_j } } \right\|^\alpha   - h_{R_l D } d_{R_l D}^{ - \alpha } \left\| {x_{R_l }  - x_{E_j } } \right\|^\alpha  } \right]dx_{E_j } }
\end{equation}
\rule{\linewidth}{1pt}\end{figure*}

Notice that  $g\left( {d_{SR_l }^\alpha  ,d_{R_l D }^\alpha  } \right)$   in (26) accounts for the statistical dependence between the received SNRs  at two different locations.

From (26), we can find that it is very difficult to obtain a close-form expression of $g\left( {d_{SR_l }^\alpha  ,d_{R_l D }^\alpha  } \right)$. However, we can get the lower bound of (25). It shows that $g\left( {d_{SR_l }^\alpha  ,d_{R_l D }^\alpha  } \right)$    is non-negative, then, the lower bound of secure connection probability can be written as (27), given at the top of the next page.
 \setcounter{equation}{26}
\begin{figure*}[bt!]
\begin{equation}\label{27}
 P_{C\_R_l \_lower}  \\= E_{h_{SR_l } ,h_{R_l D} } \left\{ {\exp \left[ { - \pi \lambda _E \Gamma \left( {1 + \frac{2}{\alpha }} \right)\left( {d_{SR_l }^2 \left( {h_{SR_l } } \right)^{ - \frac{2}{\alpha }}  + d_{R_l D}^2 \left( {h_{R_l D} } \right)^{ - \frac{2}{\alpha }} } \right)} \right]} \right\} \\
\end{equation}
\rule{\linewidth}{1pt}\end{figure*}

Substituting  $d_{\!SR_l}^{}  \!=\sqrt {r^2 + \frac{{d_{SD}^2 }}{4} \!- rd_{SD}^{} \cos \theta }$, and  $d_{R_lD}^{}=\sqrt {r^2 +\! \frac{{d_{SD}^2 }}{4} +\! rd_{SD}^{} \cos \theta }$ into (27), the lower bound of secure connection probability  can be obtained as (28), written at the top of the next page.
 \setcounter{equation}{27}
\begin{figure*}[bt!]
\begin{equation}\label{28}
P_{C\_R_l \_lower}  =\ E_{h_{SR_l ,h_{R_l D} } } \left\{{\exp \left[ { - \pi \lambda _E \Gamma \Big( {1+ \frac{2}{\alpha }} \Big)\Big( {\big( {r^2\!  + \frac{{d_{SD}^2 }}{4} - \!rd_{SD}^{} \cos \theta } \big)h_{SR_l }^{ - \frac{2}{\alpha }}  +\big( {r^2  + \!\frac{{d_{SD}^2 }}{4}+\! rd_{SD}^{} \cos \theta } \big)h_{R_l D}^{ - \frac{2}{\alpha }} } \Big)} \right]} \right\}
\end{equation}
\rule{\linewidth}{1pt}\end{figure*}

 Using Jensen$'$s inequality, (28) can be expressed as
\begin{equation}\label{29}
P_{C\_R_l \_lower} {\rm{ }} > \exp \left[{ - \frac{{Ad_{SD}^2 }}{2} - 2Ar^2 } \right]
\end{equation}

Interestingly, we find that (29) is the same as (13). Both  (28) and (29) are  lower bounds of secure  connection probability  for  non-colluding eavesdroppers, while
it is obvious that the lower bound of (28) is tighter than that of (29).

From (25) and (27), we can get (30), shown at the top of the next page.
\setcounter{equation}{29}
\begin{figure*}[bt!]
 \begin{equation}\label{30}
 \begin{array}{l}
 \frac{{P_{C\_R_l } }}{{P_{C\_R_l \_lower} }} = \frac{{E_{h_{SR_l } ,h_{R_l D} } \left\{ {\exp \left[ { - \pi \lambda _E \left( {d_{SR_l }^2 \Gamma \left( {1 + \frac{2}{\alpha }} \right)h_{SR_l }^{ - \frac{2}{\alpha }}  + d_{R_l D}^2 \Gamma \left( {1 + \frac{2}{\alpha }} \right)h_{R_l D}^{ - \frac{2}{\alpha }} } \right) + \lambda _E g\left( {d_{SE_j }^\alpha  ,d_{R_l E_j }^\alpha  } \right)} \right]} \right\}}}{{E_{h_{SR_l } ,h_{R_l D} } \left\{ {\exp \left[ { - \pi \lambda _E \left( {d_{SR_l }^2 \Gamma \left( {1 + \frac{2}{\alpha }} \right)h_{SR_l }^{ - \frac{2}{\alpha }}  + d_{R_l D}^2 \Gamma \left( {1 + \frac{2}{\alpha }} \right)h_{R_l D}^{ - \frac{2}{\alpha }} } \right)} \right]} \right\}}} \\
   ~~~~~~~~~~~~~  < \frac{{\min \left\{ {E_{h_{SR_l } } \left\{ {\exp \left[ { - \lambda _E \pi d_{SR_l }^2 \Gamma \left( {1 + \frac{2}{\alpha }} \right)h_{SR_l }^{ - \frac{2}{\alpha }} } \right]} \right\},E_{h_{R_l D} } \left\{ {\exp \left[ { - \lambda _E \pi d_{R_l D}^2 \Gamma \left( {1 + \frac{2}{\alpha }} \right)h_{R_l D}^{ - \frac{2}{\alpha }} } \right]} \right\}} \right\}}}{{E_{h_{SR_l } } \left\{ {\exp \left[ { - \pi \lambda _E d_{SR_l }^2 \Gamma \left( {1 + \frac{2}{\alpha }} \right)h_{SR_l }^{ - \frac{2}{\alpha }} } \right]} \right\}E_{h_{R_l D} } \left\{ {\exp \left[ { -\pi \lambda _E d_{R_l D}^2 \Gamma \left( {1 + \frac{2}{\alpha }} \right)h_{R_l D}^{ - \frac{2}{\alpha }} } \right]} \right\}}} \\
    ~~~~~~~~~~~~~ = \min \Bigg\{ {\frac{1}{{E_{h_{SR_l } } \left\{ {\exp \left[ { - \lambda _E \pi d_{SR_l }^2 \Gamma \left( {1 + \frac{2}{\alpha }} \right)h_{SR_l }^{ - \frac{2}{\alpha }} } \right]} \right\}}},\frac{1}{{E_{h_{R_l D} } \left\{ {\exp \left[ { - \lambda _E \pi d_{R_l D}^2 \Gamma \left( {1 + \frac{2}{\alpha }} \right)h_{R_l D}^{ - \frac{2}{\alpha }} } \right]} \right\}}}} \Bigg\} \\
 \end{array}
\end{equation}
\rule{\linewidth}{1pt}\end{figure*}

 It is shown that when $\lambda _E  \to 0$, $\frac{{P_{C\_R_l } }}{{P_{C\_R_l \_lower} }} \to 1$.
  Fig.3 shows the secure connection probability for non-colluding eavesdroppers as a function of  $\lambda _E$.
  In Fig.3, we can obtain the same results with colluding eavesdroppers. What is more, we can find that the lower bound of (28) is tighter than that of (29).



\begin{figure}
\begin{center}
  \includegraphics[width=3.4in,angle=0]{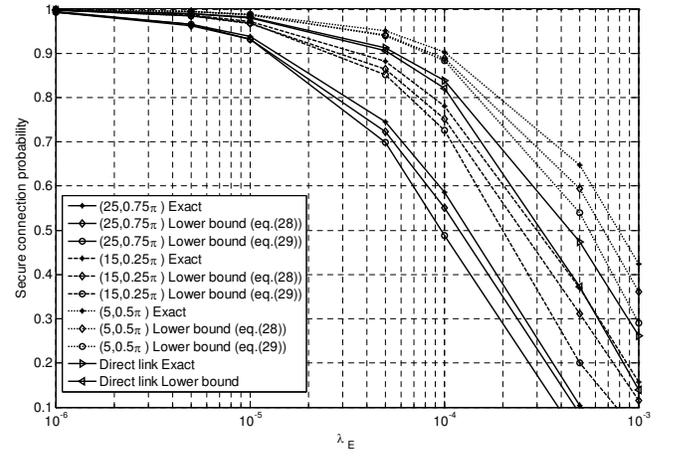}\\
  \caption{Secure connection probability for non-colluding eavesdroppers as a function of  $\lambda _E$
  when  $\alpha  = 4$, $d_{SD}  = 20m$.}
\end{center}
\end{figure}
 \subsection{Relay Selection}

We can improve the secure connection probability through a relay selection strategy. Based on (29), the optimum relay selection strategy for non-colluding eavesdroppers is still the relay which is the nearest to the midpoint between source and destination. Though  (29) is  the lower bound of non-colluding eavesdroppers, we want to obtain the tighter lower bound instead of (29).  Then,
based on (16) and (28), we can get (31), shown at the top of the next page.
\setcounter{equation}{30}
\begin{figure*}[bt!]
 \begin{equation}\label{31}
 \begin{array}{l}
 P_{C\_R_J \_lower}  = E_{h_{SR_J } ,h_{R_J D} } \left\{ {\int_{\mathbb{R}^2 } {\exp \left[ { - \pi \lambda _E \Gamma \left( {1 + \frac{2}{\alpha }} \right)\left( {h_{SR_J }^{ - \frac{2}{\alpha }}  + h_{R_J D}^{ - \frac{2}{\alpha }} } \right)\left( {r^2  + \frac{{d_{SD}^2 }}{4}} \right)} \right]} } \right. \\
   ~~~~~~~~~~~~~~~   \times \left. {\lambda _R \exp \left[ { - \pi \lambda _E \Gamma \left( {1 + \frac{2}{\alpha }} \right)\left( {h_{R_J D}^{ - \frac{2}{\alpha }}  - h_{SR_J }^{ - \frac{2}{\alpha }} } \right)rd_{SD}^{} \cos \theta  - \lambda _R \pi r^2 } \right]dx_{R_J } } \right\} \\
  ~~~~~~~~~~~~~~~  = E_{h_{SR_J } ,h_{R_J D} } \left\{ {\int_0^{2\pi } {\int_0^{ + \infty } {\exp \left[ { - \pi \lambda _E \Gamma \left( {1 + \frac{2}{\alpha }} \right)\left( {h_{SR_J }^{ - \frac{2}{\alpha }}  + h_{R_J D}^{ - \frac{2}{\alpha }} } \right)\left( {r^2  + \frac{{d_{SD}^2 }}{4}} \right)} \right]} } } \right. \\
 ~~~~~~~~~~~~~~~   \times \left. {\lambda _R \exp \left[ { - \pi \lambda _E \Gamma \left( {1 + \frac{2}{\alpha }} \right)\left( {h_{R_J D}^{ - \frac{2}{\alpha }}  - h_{SR_J }^{ - \frac{2}{\alpha }} } \right)rd_{SD}^{} \cos \theta  - \lambda _R \pi r^2 } \right]rdrd\theta } \right\} \\
 \end{array}
\end{equation}
\rule{\linewidth}{1pt}\end{figure*}

Using the fact that $I_0 \left( x \right)\mathop  = \limits^{} \frac{1}{\pi }\int_0^\pi  {e^{ \pm x\cos \theta } } d\theta $, (31) can be obtained as (32), written at the top of the next page.
\setcounter{equation}{31}
\begin{figure*}[bt!]
 \begin{equation}\label{32}
\begin{array}{l}
 P_{C\_R_J \_lower} {\rm{ }} = E_{h_{SR_J } ,h_{R_J D} } \left\{ {\int_0^{+ \infty } {\exp \left[ { - \pi \lambda _E \Gamma \left( {1 + \frac{2}{\alpha }} \right)\left( {h_{SR_J }^{ - \frac{2}{\alpha }}  + h_{R_J D}^{ - \frac{2}{\alpha }} } \right)\left( {\frac{{d_{SD}^2 }}{4} + r^2 } \right)} \right]} } \right. \\
 ~~~~~~~~~~~~~~\left. { \times 2\pi I_0 \bigg( {\pi \lambda _E \Gamma \left( {1 + \frac{2}{\alpha }} \right)\left( {h_{R_J D}^{ - \frac{2}{\alpha }}  - h_{SR_J }^{ - \frac{2}{\alpha }} } \right)rd_{SD}^{} } \bigg)\lambda _R \exp \left[ { - \lambda _R \pi r^2 } \right]rdr} \right\} \\
 \end{array}
\end{equation}
\rule{\linewidth}{1pt}\end{figure*}

Using the fact that $I_0 \left( x \right) = \sum\limits_{n = 0}^{ + \infty } {\frac{{x^{2n} }}{{\left( {n!} \right)^2 2^{2n} }}}$, (32) can be derived as (33), given at the top of the next page.
\setcounter{equation}{32}
\begin{figure*}[bt!]
\begin{equation}\label{33}
\begin{array}{l}
P_{C\_R_J \_lower} {\rm{ }} = E_{h_{SR_J } ,h_{R_J D} } \left\{ {\sum\limits_{n = 0}^{ + \infty } {\bigg( {\pi \lambda _E \Gamma \left( {1 + \frac{2}{\alpha }} \right)\left({h_{R_J D}^{ - \frac{2}{\alpha }}  - h_{SR_J }^{ - \frac{2}{\alpha }} } \right)d_{SD}^{} } \bigg)^{2n} \frac{{2\pi \lambda _R }}{{\left( {n!} \right)^2 2^{2n} }}} } \right. \\
~~~~~~~~~~~~~~~ \left. { \times \int_0^{+ \infty } {\exp \bigg[{ - \pi \lambda _E \Gamma \left( {1 + \frac{2}{\alpha }} \right)\left( {h_{SR_J }^{ - \frac{2}{\alpha }}  + h_{R_J D}^{ - \frac{2}{\alpha }} } \right)\left( {\frac{{d_{SD}^2 }}{4} + r^2 } \right) - \lambda _R \pi r^2 } \bigg]} r^{2n + 1} dr} \right\} \\
 \end{array}
\end{equation}
\rule{\linewidth}{1pt}\end{figure*}

Using  [20,3.381.10], (33) can be written as (34), derived at the top of the next page.
\setcounter{equation}{33}
\begin{figure*}[bt!]
\begin{equation}\label{34}
P_{C\_R_J \_lower} {\rm{ }}\!
=\! E_{h_{SR_J } ,h_{R_J D} }\!\! \left\{ {\pi \sum\limits_{n = 0}^{ + \infty }\!\!{\frac{\lambda _R {\bigg( \!\!\!{\pi \lambda _E \Gamma \left( {1 + \frac{2}{\alpha }} \right)\left( {h_{R_J D}^{ - \frac{2}{\alpha }}  - h_{SR_J }^{ - \frac{2}{\alpha }} } \right)d_{SD}^{} } \bigg)^{2n}\!\!\! \exp \left[ { - \pi \lambda _E \Gamma \left( {1 +  \frac{2}{\alpha }} \right)\left( {h_{SR_J }^{ - \frac{2}{\alpha }}  + h_{R_J D}^{ - \frac{2}{\alpha }} } \right)\frac{{d_{SD}^2 }}{4}} \right]}}{{n!2^{2n} \left( {\pi \lambda _E \Gamma \left( {1 + \frac{2}{\alpha }} \right)\big( {h_{SR_J }^{ - \frac{2}{\alpha }}  + h_{R_J D}^{ - \frac{2}{\alpha }} } \big)}\! \right)^{1 + n} }}} }\! \right\}
\end{equation}
\rule{\linewidth}{1pt}\end{figure*}


Fig.4 shows the secure connection probability as a function of  $\lambda _E$ when  $\lambda _R  = 10^{ - 2} , 10^{ - 3} $, $\alpha  = 4$ and  $d_{SD}  = 20$. Firstly, we can find that our theoretical results for direct transmission match the simulation results perfectly for two strategies. The lower bounds of  secure connection  probability for relay transmission for both the two strategies are very close to the simulation results, when $\lambda _E$ is low, and the gap between them becomes larger with $\lambda _E$ increasing. Secondly,  the secure connection  probability for non-colluding eavesdroppers is better than that of  colluding eavesdroppers. Thirdly, it is shown that the secure connection  probability for relay transmission depends on the relay density. When  $\lambda _R$ is low, the secure connection  probability for relay transmission is worse. From Fig. 4, we do not know how to decide whether or not to cooperate distinctly.
\begin{figure}
\begin{center}
  \includegraphics[width=3.4in,angle=0]{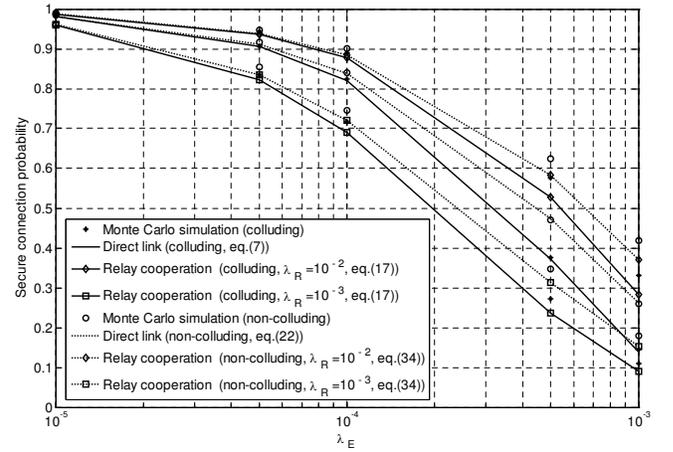}\\
  \caption{Secure connection probability  as a function of  $\lambda _E$ when $\lambda _R  = 10^{ - 2} , 10^{ - 3} $, $\alpha  = 4$,  and  $d_{SD}  = 20m$}
\end{center}
\end{figure}

\section{The Condition for Relay Transmission}

 \subsection{Colluding eavesdroppers}

The relay transmission is not necessary if the direct transmission is strong enough. According to $d_{SD}$, we can know how strong the direct transmission is.
For colluding eavesdroppers, to avoid relaying for the users having strong direct transmission we define a target secure connection  probability $\delta$ constraint for direct transmission, thus the upper bound $d_{SD\_D}$  for $d_{SD} $  based on a target secure connection  probability $\delta$ can be written as
\setcounter{equation}{34}
\begin{equation}\label{35}
\begin{array}{l}
 d_{SD\_D} \left( \delta  \right) = \max \left\{ {d_{SD} :\exp \left( { - Ad_{SD}^2 } \right) > \delta } \right\} \\
 ~~~~~~~~~~~~ = \sqrt {\frac{{\ln \left( {{1 \mathord{\left/
 {\vphantom {1 \delta }} \right.
 \kern-\nulldelimiterspace} \delta }} \right)}}{A}}  \\
 \end{array}
\end{equation}

Similar to the direct transmission, the upper bound $d_{SD\_R}$  for  $d_{SD}$ using relay transmission based on a target secure connection  probability $\delta$ can be written as
\begin{equation}\label{36}
\begin{array}{l}
 d_{SD\_R} \left( \delta  \right) = \max \left\{ {d_{SD} :\frac{{\lambda _R \pi }}{{2A + \lambda _R \pi }}\exp \left( { - A\frac{{d_{SD}^2 }}{2}} \right) > \delta } \right\} \\
 ~~~~~~~~~~~~ = \sqrt {\frac{2}{A}\ln \left( {\frac{{\delta \left( {2A + \lambda _R \pi } \right)}}{{\lambda _R \pi }}} \right)^{ - 1} }  \\
 \end{array}
 \end{equation}

If  $\frac{{\lambda _R \pi }}{{\delta \left( {2A + \lambda _R \pi } \right)}} < 1$, we define  $d_{SD\_R} \left( \delta  \right) = 0$, which means relay transmission strategy does not work for colluding eavesdroppers.
 From (35), we can know that the target secure connection probability can not be satisfied using direct transmission, when  $d_{SD}  > d_{SD\_D}$.
When $d_{SD}  > \sqrt {\frac{{\ln \left( {{1 \mathord{\left/
 {\vphantom {1 \delta }} \right.
 \kern-\nulldelimiterspace} \delta }} \right)}}{A}} $, we want to find how high $\lambda _R $    can satisfy the target secure connection probability for relay transmission.

 Then, we define  the secure gain  $ G\left( \delta  \right)$ as the ratio of the farther distance  achieved  between the source and the destination by using relay transmission and direct transmission, and  the secure gain for colluding eavesdroppers is shown as
\begin{equation}\label{37}
G\left( \delta  \right) = \sqrt {\frac{{2\ln \left( {\frac{{\lambda _R \pi }}{{\delta \left( {2A + \lambda _R \pi } \right)}}} \right)}}{{\ln \left( {{1 \mathord{\left/
 {\vphantom {1 \delta }} \right.
 \kern-\nulldelimiterspace} \delta }} \right)}}}
 \end{equation}

From (37), we can get when $\lambda _R  > \frac{{2A\left( {\delta  + \sqrt \delta  } \right)}}{{\pi \left( {1 - \delta } \right)}}$, $ G\left( \delta  \right)>1$.
Because the source knows the relative distances between the destination, the relays, and itself, it can decide whether a relay is needed through (35), (36) and (37). The detailed procedure is listed as follows:
 when $d_{SD}  < d_{SD\_D}$ , the direct transmission is strong enough for the target secure connection  probability, then relay transmission is not necessary. Secondly, when  $d_{SD}  > d_{SD\_D}$, the direct transmission can not satisfy the target secure connection  probability, then relay transmission can be used if $\lambda _R  > \frac{{2A\left( {\delta  + \sqrt \delta  } \right)}}{{\pi \left( {1 - \delta } \right)}}$. Then, the relay selection strategy is based on (15).

 \subsection{Non-colluding eavesdroppers}
For non-colluding eavesdroppers, we find that (22) and (34) are not easy to analyze because of  the integrals. However, we can also find that (7) and (17) is  the lower bound for direct transmission and relay transmission of colluding eavesdroppers, respectively. Then, we can use the results of (35) and (36) to analyze whether or not to relay transmission  for non-colluding eavesdroppers. Because (7) is the lower bound for direct transmission of colluding eavesdroppers, thus when $d_{SD}  > \sqrt {\frac{{\ln \left( {{1 \mathord{\left/
 {\vphantom {1 \delta }} \right.
 \kern-\nulldelimiterspace} \delta }} \right)}}{A}} $, the given target secure connection  probability $\delta$ can  be satisfied using direct transmission, while if $\lambda _R  > \frac{{2A\left( {\delta  + \sqrt \delta  } \right)}}{{\pi \left( {1 - \delta } \right)}}$, the given target secure connection  probability $\delta$ must  be satisfied using relay transmission, then we can regard the results of (35) and (36) as a sufficient condition for using relay transmission  for non-colluding eavesdroppers. Thus, the detailed procedures for determining whether a relay is needed and which relay is selected are the same as the ones in the colluding eavesdroppers case.

Fig. 5 shows the secure gain as a function of $\lambda _R$  when   $\alpha  = 4$, and  $\lambda _E  = 10^{ - 3} ,10^{ - 5}$. From (36), we can find that the secure gain become better with the increase of  $\lambda _R$. When  $\lambda _R \to + \infty$,  $G\left( \delta  \right) \to \sqrt 2$. When  $G\left( \delta  \right) < 1$, it means that the secure connection  probability for direct transmission is better than that of relay transmission. In this case, relay transmission is not necessary even if the  target secure connection  probability can not be satisfied using direct transmission.
When  $G\left( \delta  \right) > 1$, it means that the  $d_{SD\_R}$ is farther than  $d_{SD\_D}$. When  $d_{SD\_D}  > d_{SD}$, the direct transmission is still  strong enough for the target secure connection  probability, then relay transmission is not necessary. Only when  $d_{SD\_D}  < d_{SD}$, relay transmission is necessary. From Fig.5, we can know how high $\lambda _R$ can match the case.

Fig. 6 shows the secure connection  probability as a function of $\lambda _E$  when  $\alpha  = 4$,  $\lambda _R  = 10^{ - 3} ,10^{ - 4}$, $\lambda _E  = 10^{ - 5}$,  and  $\delta  = 0.7$.
From (35), the upper bound $d_{SD\_D}$  for $d_{SD} $ is 85m.
  From (36), the upper bound $d_{SD\_R}$  for $d_{SD} $ is 115m and 58m  at  $\lambda _R  = 10^{ - 3} ,10^{ - 4}$, respectively.
 In Fig. 6, for colluding eavesdroppers, we can find that when  $d_{SD}  > 85$m, direct transmission can not satisfy the target secure connection  probability. However, only if  $\lambda _R  > \frac{{2A\left( {\delta  + \sqrt \delta  } \right)}}{{\pi \left( {1 - \delta } \right)}} = {\rm{1}}{\rm{.6}} \times {\rm{10}}^{ - 4}$, the secure connection  probability for relay transmission can satisfy the target secure connection  probability.
 Then, it is shown that when $\lambda _R  =10^{ - 4}<{\rm{1}}{\rm{.6}} \times {\rm{10}}^{ - 4}$, relay transmission can not satisfy the target secure connection  probability, while when $\lambda _R  =10^{ - 3}>{\rm{1}}{\rm{.6}} \times {\rm{10}}^{ - 4}$, the target secure connection  probability can be satisfied using relay transmission.
For non-colluding eavesdroppers, because (35) is the lower bound for direct transmission, we can  find that when  $d_{SD}  > 85$m, direct transmission can still satisfy the target secure connection  probability.
 However, when $\lambda _R >{\rm{1}}{\rm{.6}} \times {\rm{10}}^{ - 4}$, the secure connection  probability for relay transmission must satisfy the target secure connection  probability. Thus,
  the analysis results of (35) and (36) are  regarded as a sufficient condition for using relay transmission in non-colluding eavesdroppers case.

\begin{figure}
\begin{center}
  \includegraphics[width=3.4in,angle=0]{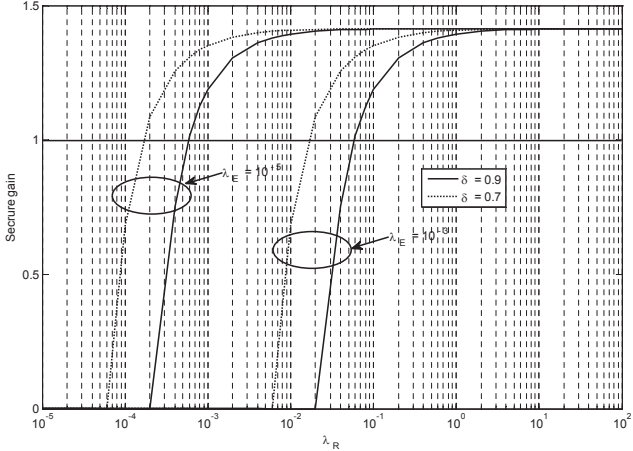}\\
  \caption{Secure gain as a function of $\lambda _R$  when   $\alpha  = 4$, and  $\lambda _E  = 10^{ - 3} ,10^{ - 5}$.}
\end{center}
\end{figure}

\begin{figure}
\begin{center}
  \includegraphics[width=3.4in,angle=0]{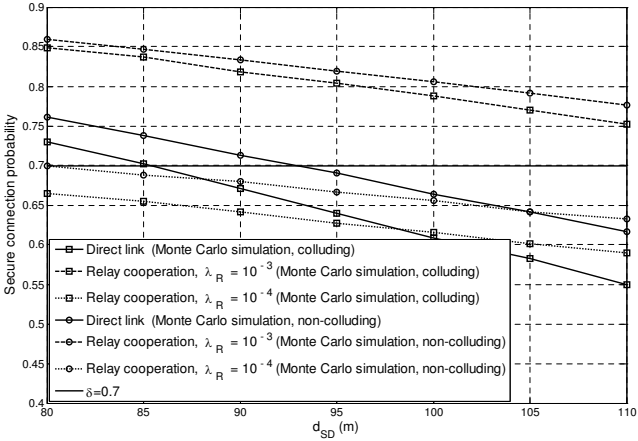}\\
  \caption{Secure connection probability as a function of $\lambda _E$  when  $\alpha  = 4$,  $\lambda _R  = 10^{ - 3} ,10^{ - 4}$,  $\lambda _E  = 10^{ - 5}$,  and  $\delta  = 0.7$.}
\end{center}
\end{figure}

\section{Conclusions}
In this paper, we have analyzed the secure connection probability of direct transmission and relay transmission for colluding eavesdroppers and non-colluding eavesdroppers strategies, where the distributions of relays and eavesdroppers follow homogenous PPPs. The lower bound expressions of secure connection probability using RaF for colluding eavesdroppers and non-colluding eavesdroppers strategies are  obtained, and it shows that the lower bound gives accurate approximation of the exact performance in the small eavesdropper density regime.  Comparing the direct transmission with the relay transmission, we find that whether or not to relay transmission depends on the relay density and the distance between the source and destination for  a given target secure connection  probability. The results obtained from this study provide useful design insights for relay networks with security constraints. This paper focuses on the selection of one relay (i.e. best relay). An interesting future work is to consider multi-relay transmission and determine the condition under which relaying gives more secure connection.


%

\ifCLASSOPTIONcaptionsoff
  \newpage
\fi

\end{document}